# Smart Tracking Tray System for A Smart and Sustainable Wet Lab Community

Nan XU[1], *Graduate Student Member, IEEE*, Jingchen LI[2,3], Yue YU[2,3], Yang LI[4], Jinglei YANG[5,6,*]

[1] nxuab@connect.ust.hk Division of Emerging Interdisciplinary Areas, Hong Kong University of Science and Technology, HKSAR, China
[2] jlieg / yyubv@connect.ust.hk Division of Social Science, Hong Kong University of Science and Technology, HKSAR, China
[3] jlieg / yyubv@connect.ust.hk Department of Computer Science, Hong Kong University of Science and Technology, HKSAR, China
[4] ylikp@connect.ust.hk Artificial Intelligence Thrust, Hong Kong University of Science and Technology (Guangzhou), China
[5] maeyang@ust.hk Department of Mechanical & Aerospace Engineering, Hong Kong University of Science and Technology, HKSAR, China
[6] maeyang@ust.hk Sustainable Energy and Environment Thrust, Hong Kong University of Science and Technology (Guangzhou), China

## I. Background

The laboratories and research institutes are the major place of doing cutting-edge scientific exploration. Hundreds of millions of research papers [1] were formed from front-line labs. Behind this glorious achievement were three unsustainable facts:

1. The reproducibility of the experiment is low. The replication is difficult and time-consuming. Besides, error tracking is complex, with unexpected accidents occurring.

2. The experimental data are too costly to meet data analysis needs. Meanwhile, many precious failed experiments have not been effectively explored, analyzed but are directly ignored or discarded.

3. The global experimental innovation efficiency has encountered a bottleneck. Moore's Law has doubled, and the experimental personnel invested must be multiplied by 18 times [2].

The above facts increasingly require more human investment in innovative experimental design and analysis of results. However, the laboratory operating environment has not been subversively transformed for centuries. The front-end researchers are majorly trapped by chemical/vessel finding, experimenting, and other low-value-added operations. Therefore, digital transformation of laboratories with enabling technologies is essential and intrinsic to free researchers' productivity, and establishing a sustainable & smart research community. In China, only about 1/3 of university labs are equipped with lab management software. The remaining 2/3 relies on Excel or manual recording methods, where 88% of spreadsheets contain incorrect information, leading to potential accidents [3]. One U.S. research [4] shows that around 40% of lab incidents are related to improper chemical storage and handling. Besides, a significant amount of expired or rarely used chemicals are wasted, posing a financial and environmental burden. These chemicals, however, can be better exploited by a chemical exchange platform.

With the development of Information Technology, various patents and papers related to the laboratory's management [5,6] have emerged. Based on the functionalities, they can be categorized into storage management and supply chain, fund application, lab automation, and lab notebook [7-9]. However, most products merely digitalize the records without further innovations to help manage the lab smartly. Also, literature covering hardware, equipment, and automatic machines mainly focuses on the logistics industry and inventory management [10,11]. It is not compatible with real wet lab scenarios due to the lack of market research. Therefore, many individual labs have tried the products with interest and then returned to traditional labor-intensive management.

This abstract proposed a smart tracking system to track the chemicals in an automatic and timely approach.

## II. Proposed Solution

### A. System design

The system consists of three layers. As shown in Fig. 1, the hardware block is the most fundamental part of the system. It includes RFID readers and weight sensors and can be extended with other functional sensors. They can detect the chemicals' information, corresponding historical weight changes, and related user information. The real-time data acquired will be sent to the central database in the software backend, for further alignment and processing. A user-friendly interface will present clients with comprehensive chemical information through cross-platform software.

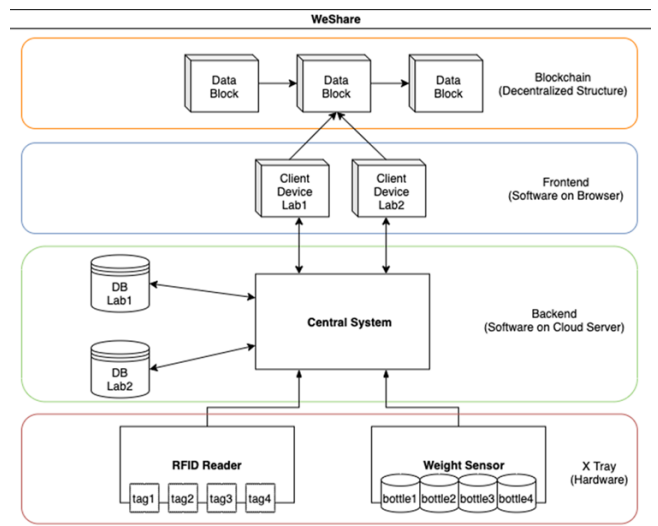

Fig. 1 The IoT system of the Smart Tracking Tray.

### B. Hardware Modules and Integration

Fig. 2 demonstrates the hardware part: X-Tray. The left picture shows that Arduino 33 IoT MCU combines an RFID module, Weight sensing load cells, and other peripheral modules. Strain gauge-type transducer weight sensors are settled with the HX711 ADC module. JRD-100 ultra-high frequency (UHF) active RFID module is for detection. The customized Printed Circuit Board (PCB) integrates these functional parts. The photo on the right shows an actual application of an X-Tray in the chemical storage cabinet.

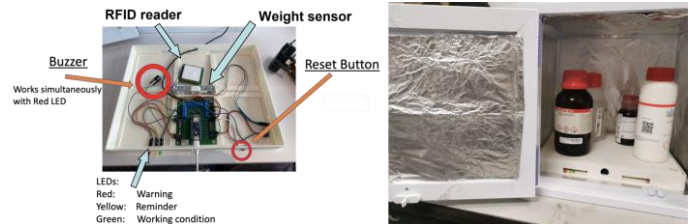

Fig. 2 The hardware system and application in the chemical storage cabinet.

Fig. 3 illustrates the operational flow of the hardware system. With the pre-attached RFID Tags, this hardware system can detect the chemicals and corresponding weight difference after each take-return action. The captured data will be sent to the cloud server via a Wi-Fi module.

### C. Software and User Interface

Fig. 4 shows the data processing that takes place in the backend.

When the background detects that the weight is not stable, it considers that an operation event occurred. And the system internally marks the X-Tray with pending status, records the weight of the X-Tray, from the system and the presence tag I.D., and waits for the weight to stabilize. While waiting, all tag I.D.s are recorded and used for match containers.

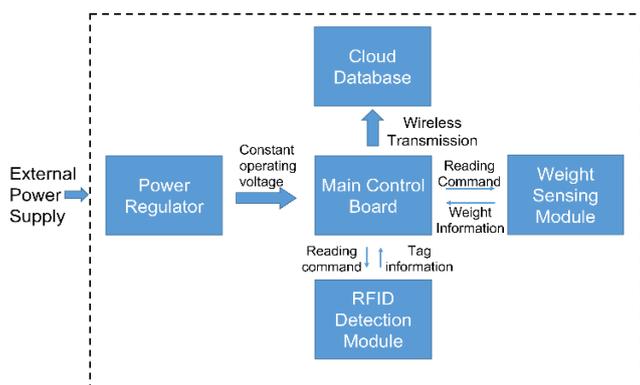

Fig. 3 System design of the RFID and Weight Sensor Combination.

Once the weight is stable, the system calculates the container weight change using the difference between the current stable weight and the weight before the event. Both the tag I.D. recorded during and before the event are used to match the operation's container.

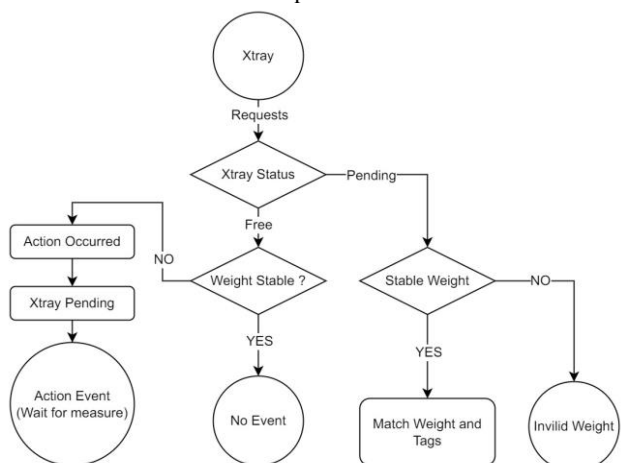

Fig. 4 The flow chart of data processing and matching.

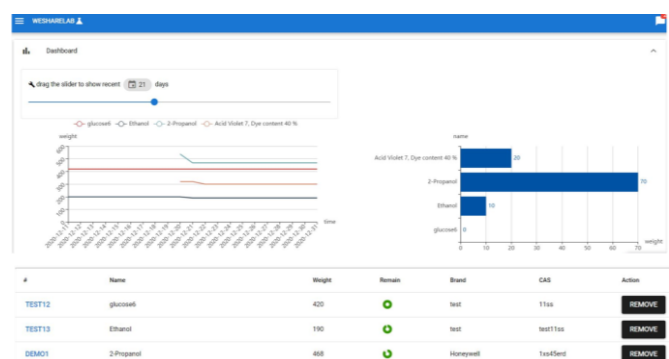

Fig. 5 The software platform: chemical index view.
(http://demo.wesharetechnology.com/#/)

The software interface (Fig. 5) can cater to the needs of different lab personnel. For example, an experimenter may check the availability of a chemical through the system and know its remaining quantity and location. Lab managers can see the chemical's consumption history and trends through the dashboard. The interactive and intuitive visualizations enable the lab manager to grasp the inventory status at a glance. The demo website can be found at http://demo.wesharetechnology.com/#/.

Apart from the basic features, we hope to utilize the usage data for prediction and pattern-finding. By rough estimation, a few hundred chemical data sets will be generated for a Ph.D. student who works on the chemical-related research topic. A moderate-size lab typically has 5-10 Ph.D. students generating data days and nights. One can imagine the size of the database accumulated over the years. The voluminous data has tremendous potential in artificial intelligence-guided big data modeling and is extremely valuable in uncovering the hidden patterns in experiments. To be more concrete, the system may remind lab managers to supplement the chemicals before they run out according to prediction; it may also give suggestions to researchers on how to adjust their formula.

In addition to the comprehensive visualization, some essential data and records from different labs must be safely stored on a trustable platform with adequate security assurance. Besides, if there is any dispute about academic research's experiment results, we need to trace back to find the evidence. Thus, we will enable users to upload their lab notes or experiment results to public blockchains(Ethereum, etc.) to ensure authenticity and originality. These proofs can be got and verified anywhere in the world and do not need any additional trust.

## III. CONCLUSION

In this abstract, we proposed a Smart Tracking Tray system for chemical management. We have already received positive feedback from pilot tests in several labs at HKUST. The system benefits various lab users in their daily work and improves their working efficiency. In the long run, we believe it will play an essential role in promoting the efficient use of lab resources and achieving the goal of sustainable labs.